# Using statistical control charts to monitor duration-based performance of project

Nooshin Yousefi[a], Ahmad Sobhani[b], Leila Moslemi Naeni[c], Kenneth R. Currie[d]

[a] Department of Industrial & Systems Engineering, Rutgers University, Piscataway, NJ, 08854
[b] Department of decision and Information sciences, scholl of business administration Oakland University
[c] School of Built Environment, University of Technology Sydney, Australia
[d] Department of Industrial Engineering, West Virginia University, Morgantown, WV,26505

**Abstract**

Monitoring of project performance is a crucial task of project managers that significantly affect the project success or failure. Earned Value Management (EVM) is a well-known tool to evaluate project performance and effective technique for identifying delays and proposing appropriate corrective actions. The original EVM analysis is a monetary-based method and it can be misleading in the evaluation of the project schedule performance and estimation of the project duration. Earned Duration Management (EDM) is a more recent method which introduces metrics for the project schedule performance evaluation and improves EVM analysis. In this paper, we apply statistical control charts on EDM indices to better investigate the variations of project schedule performance. Control charts are decision support tools to detect the out of control performance. Usually project performance measurements are auto-correlated and not following the normal distribution. Hence, in this paper, a two-step adjustment framework is proposed to make the control charts applicable to non-normal and auto-correlated measurements. The case study project illustrates how the new method can be implemented in practice. The numerical results conclude that that employing control chart method along with analyzing the actual values of EDM indices increase the capability of project management teams to detect cost and schedule problems on time.

**Keywords:** Earned Duration Management, Statistical Control Charts, Project Performance Evaluation.

1. ## Introduction

Project management is the application of processes, methods, knowledge, skills, and experience to achieve project objectives and minimize the risk of project failure (Meredith, Mantel, & Shafer 2015). With respect to the complexity of today's projects, many project management tools are developed to support managers in controlling and aligning projects with their planned timelines, scopes and budgets (Project Management Institute, 2008). One of the important project management methods is Earned Value Management (EVM) that measures the performance and progress of a project in an objective manner by integrating scope, schedule and cost elements of the project (Project Management Institute, 2005; Salehipour, Naeni, Khanbabaei, & Javaheri 2015). The Project Management Institute (PMI) extensively discussed the basic terminology and formulas of the EVM method in 2000. This terminology was simplified, and more details on EVM were provided in the 3$^{rd}$ edition of Project Management Body Of Knowledge (PMBOK) (Project Management Institute, 2004). EVM mainly focuses on the accurate measurement of the work in progress against a detailed plan of the project by creating performance indices for cost, time and project completion (Fleming & Koppelman, 1996; Chen, Chen, & Lin,2016). Comparing the planned values of these performance indices (determined according to the baseline plan) with their actual records, will enable managers to keep the project on the track and figure out how will be the project future performance (Project management Institute, 2013).

There has been some research discussed the application of EVM methodology and associated benefits to control/monitor projects in different disciplines (e.g., Cioffo, 2006; Vandevoorde & Vanhoucke, 2006; Abba & Niel, 2010; Turner, 2010). For instance, Rodríguez et al., (2017) discussed the benefits of applying EVM for managing large projects in the aerospace industry. Batselier & Vanhoucke (2015a) used empirical data from 51 real-life projects to evaluate the accuracy of EVM time and cost indicators. Several studies have also challenged the accuracy of EVM in predicting the progress of a project and proposed extensions for the EVM methodology to provide more accurate results (e.g., Lipke, 2003; Henderson, 2003; Vandevoorde & Vanhoucke, 2006; Lipke, Zwikael, Henderson, & Anbari, 2009; Azman, Abdul-Samad, & Ismail 2013; Colin & Vanhoucke, 2014; Narbaev & Marco, 2015, Salari, Yousefi & Asgary 2016).

Lipke (2003) argued the shortcomings of the schedule indicators of EVM and their inaccurate estimated values when projects are close to their ends and their performances are poor. To improve the accuracy of EVM schedule performance indicators, Lipke introduced the concept of Earned Schedule (ES). ES is a straightforward translation of EV into time units by determining when the EV should have been earned in the baseline. Although this extension improves the accuracy of indices and project forecasts but still include the major drawbacks of EVM by considering monetary-based factors in evaluating the project schedule performance. EVM method uses monetary based factors as proxies to estimate the schedule/duration performance of a project through its life cycle. However, the employment of cost-based metrics to control the duration/schedule of a project is misleading as cost and schedule profiles are not generally the same during the project life cycle (Khamooshi & Golafshani, 2014).

Although EDM method has improved the accuracy of managers' prediction about the schedule performance of projects, there are no studies consider EDM indices to evaluate the deviations of a project from the baseline plan throughout the project life cycle. Due to several reasons such as unavailability of the resources and delays in completing tasks, the project may experience the schedule and budget overruns. Monitoring the project, detecting cost and schedule performance variations, and accomplishing corrective actions are necessary to lead a project management team towards the successful completion of the project.

One of the innovative approaches to monitoring performance deviations of a project from the base plan is to use methods that statistically track the trends of performance indices of projects (e.g., Barraza & Bueno, 2007; Aliverdi, Naeni, & Salehipour, 2013). These statistical methods will let practitioners know the acceptable range of project performance deviations. They also help project management practitioners to figure out the cause of unacceptable (non-random) deviations which may generate unexpected costs and delays in the future. Using statistical methods to analyse project data also reveals informative insights to project performance trends, especially in high priority projects (Aliverdi, Naeni, & Salehipour, 2013). This information helps project managers better understand the tendency and the direction of a project's performance in advance. Hence, these methods can minimize the project's overtime and over budget in a right time during the project life. Several studies applied statistical models to manage project performance deviations by controlling EVM indices throughout the project life cycle (e.g., Lipke, & Vaughn, 2000, Leu & Lin 2008; Aliverdi, Naeni, & Salehipour, 2013).

This paper aims to extend and improve the capability of traditional project management tools in dealing with project performance variations from the baseline plan with respect to EDM methodology that decouples schedule and cost dimensions of projects. Shewhart individuals control charts are generated individually for EDM cost and schedule performance indices to detect the presence of unacceptable (non-random) variations. According to the theory, individuals control charts are valid when the distribution of data is normal and there is no dependency/autocorrelation between them. However, these assumptions are not always valid in real-world projects. To overcome these limitations and achieve more reliable results, this study proposes a two-step adjustment framework that checks the normality data and normalizes the distribution of on-normal measurements. Also, the proposed framework also introduces the application of Autoregressive Integrated Moving Average models to remove autocorrelation effects. The results of this research demonstrate the employment of a more accurate project performance factors in controlling the deviations of a project from its baseline plan.

The remainder of the paper is organized as follows. First, an overview of EDM is provided. Second, statistical process control method and the normality and independency assumptions in using statistical control charts technique are discussed. Our proposed adjustment framework that lets users apply control charts technique on non-normal and auto correlated data sets is also introduced. Third, the numerical results of evaluating a project's cost and schedule performance deviations based on EDM measurements, estimated from a construction project case study are presented. Finally, the conclusion of the paper and the future line of research are provided.

## 2. Overview of Earned Duration Management (EDM)

Khamooshi and Golafshani (2014) discussed the deficiencies of EVM and Earned Schedule indices and introduced EDM. This method aims to decouple the cost and schedule dimensions completely. Therefore, schedule performance measures are all developed according to duration-based factors and are now accurate enough to be used in monitoring and predicting the duration trend of a project. While EVM is evaluating project progress based on comparing PV (planned value) and AC (actual cost) with EV (earned value), in EDM methodology introduced three new concepts: 1) Total Planned Duration or TPD, 2) Total Actual Duration or TAD and 3) Total Earned Duration or TED. Figure 1 illustrates the EDM model as proposed in (Khamooshi & Golafshani, 2014).

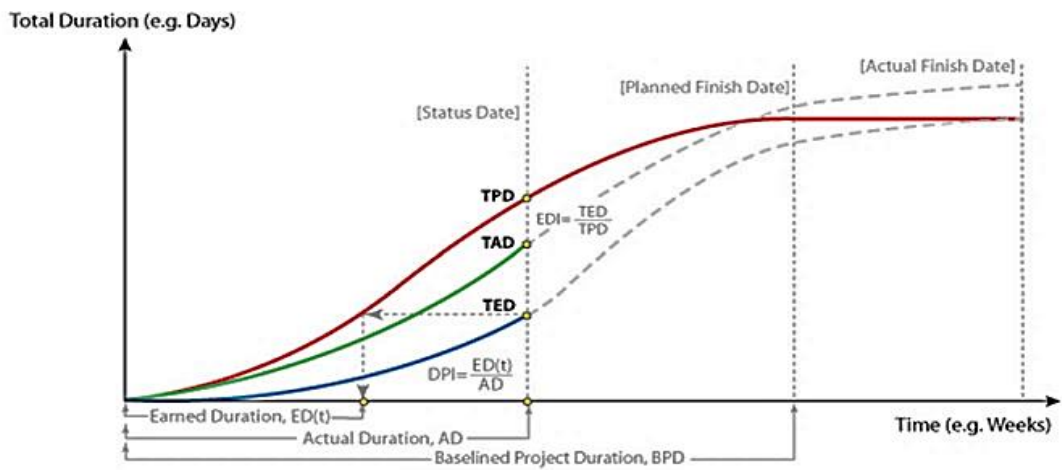

Figure 1: Conceptual EDM graph (Khamooshi & Golafshani, 2014)

TPD is a substitute for PV and refers to the summation of Planned Duration (PD) of all activities of the project. Similarly, TED refers to the summation of Earned Duration (ED) of all activities and does the same job as EV in the EVM analysis ($TED = \sum_i ED_i$ , i = activity index). TPD is also calculated according to the summation of all Planned Duration (PD) and of the same activities ($TPD = \sum_i PD_i$ , $i =$ activity index).

In EDM terminology, a schedule performance index is introduced as Earned Duration Index (EDI) which is defined as follows.

$$EDI = \frac{TED}{TPD} \qquad (1)$$

*EDI* is defined as a duration-based measurement indicator that specifies the overall work performed in terms of the duration aspect of the project by comparing it with the work planned up to any given point in time.

In EVM system, Earned Schedule (ES) is calculated by projecting the EV curve on to the PV curve. Similarly, the TED can be projected on TPD which shows the ED(t), as shown in Fig. 1. ED(t) identifies the time that achieving the current ED was expected to occur. Using ED(t) and Actual Duration (AD) a new schedule index is introduced which is called Duration Performance Index (DPI) and calculated as follows.

$$DPI = \frac{ED(t)}{AD} \qquad (2)$$

*DPI* shows how well the project is performing to meet the final completion date from the schedule performance points of view. It provides a more accurate result than $SPI_t$ and is usually compared with 1. If $DPI > 1$, it indicates a good schedule performance for the project. Hence, the project is ahead of its planned schedule. If $DPI < 1$, it indicates poor schedule performance for the project and it may behind the planned schedule. If $DPI = 1$, it shows that the project is on the plan and it does not have any delays with the respect of the planned schedule.

Using duration-based schedule performance indices enables the EDM system to provide more precisions for managers in evaluating and controlling the project from both time (schedule) and cost points of view (Khamooshi & Golafshani, 2014, Yousefi 2018). EDM indices will be used in our proposed project management control approach to track and analyze the deviations of a given project from the scheduled plan during its life cycle.

In this paper, EDI and DPI together with CPI (cost performance index which is defined as EV/AC) will be used in the proposed project management control approach to track and analyse the deviations of a given project from the scheduled plan during its life cycle.

## 3. Evaluating the Behaviour of Project Performance by Applying Individuals Control Charts

Cost and schedule indices of EDM are measured based on what has been achieved and what was going to be attained. The deviations of these performance measurement indicators would be the signs of an improper progress for the project. Generally speaking, deviations of a project's performance from the baseline plan can be clustered as controlled and uncontrolled deviations (e.g., Thor, Lundberg, Ask, Olsson, Carli, Härenstam, & Brommels 2007). Controlled deviations that are also called random or acceptable deviations, are mostly inherent in a project and are not identified very important. In contrast, the uncontrolled deviations that are also called unacceptable or non-random are often due to some deficiencies in the progress of the project that have not been noticed in the baseline plan. These unacceptable deficiencies may result in unexpected delays or over budget runs. To prevent or reduce these undesirable time and cost losses, it is necessary to monitor the behaviour of the project performance by analyzing the deviations of EDM indices during the project life cycle. The Shewhart individual statistical control charts technique is an analytical method that provides valuable information to distinguish between controlled (acceptable) and uncontrolled (unacceptable) deviations in EDM indices. This information supports project manager decisions on how to keep the project on track by detecting deficiencies in the progress of the project and taking corrective actions on time.

In the next section, we review Shewhart individual control charts technique applied in this paper to control the deviations of EDM indices over time. We also discuss the limitations of this technique and introduce our two-step adjustment framework, which overcomes these limitations and enhances the validity and practicality of the Shewhart control charts application in real world projects.

## 4. Individual Statistical Control Charts

Statistical Process Control (SPC) is a branch of statistics, which is used to monitor and control processes. Applying SPC helps managers to detect performance variations through a project life cycle and identify important factors affecting the project (Oakland, 2007). Many types of SPC tools exist for controlling, but statistical control charts technique is one of the most operational methods among them.

Shewhart proposed the fundamentals of statistical control charts in the 1920's. Although some other researchers tried to improve statistical control charts, Shewhart control charts technique is still the most useful and accurate method (Oakland, 2007). The Shewhart control charts can divide variations into two types (also called type A and B here) in a way to distinguish between sources of variations (Montgomery, 2009). Type A variations are due to chance and random causes, resulted from inherent features of the process. These variations are inevitable and also called acceptable variations. Type B variations are taken place by assignable and special causes, resulting from a problem such as human and/or machine errors (e.g., Russo, Camargo, & Fabris 2012). These variations are also called unacceptable variations. In reality, it is important to detect assignable causes that may result time and financial losses for projects. The Shewhart control charts technique is widely used because of the simplicity of the method to implement. This technique has also a high accuracy to detect even small deviations and associated sources (Franco et al., 2014). For the purpose of this study, we applied Shewhart individuals control charts to monitor measurements associated with CPI, EDI, and DPI separately.

In controlling a project performance, $Y_t$ is defined as the value (measurement) of a given performance index (e.g., CPI or EDI) in time t (t =1, 2, …,T). By tracking the behavior of $Y_t$ over a given time period, Shewhart individuals control charts are able to determine the out of control or unacceptable status due to special causes. Equations (3) to (5) demonstrate the boundaries that limit the acceptable variations of $Y_t$ over the time. UCL and LCL refer to upper control limit and lower control limit, respectively. Any values above or below these boundaries alarms that the deviation of the project performance is unacceptable due to some problems during the project progression. This warning helps project team to figure out the reasons of this unacceptable situation, and also to prevent from possible time and cost losses.

$$UCL = \bar{Y} + 3(\frac{\sigma}{\sqrt{n}}) \tag{3}$$

$$CL = \bar{Y} = \frac{\sum_{j=1}^{n} Y_j}{n} \tag{4}$$

$$LCL = \bar{Y} - 3(\frac{\sigma}{\sqrt{n}}) \tag{5}$$

$\bar{Y}$ is the average of an EDM index's measurements, where n is the total number of samples (measurements), which are taken over the time; $\sigma$ is the standard deviation of the given performance index values (collected measurements).

In order to estimate the control limits used for monitoring the progress of a project, we need an initial set of samples. These samples are used to estimate the initial control limits (Montgomery, 2009; Salehipour, Naeni, Khanbabaei, & Javaheri 2015). Then, we check the status of samples and remove the out of control ones, if assignable causes were found. New control limits are estimated after removing the out of control samples. Again, the status of the remained samples will be checked. If any out of control samples exist, they will be removed and the new control limits will be determined according to the remained samples. This procedure is repeated until all samples are inside control limits (reasoning behind repeating this procedure is well explained by Montgomery, 2009). From this point, the final control limit estimations can be used to monitor the deviation of samples (measurements). According to Xie, Goh and Kuralmani, (2002), Shewhart individuals control charts provide valid results only if measurements evaluated by this technique satisfy the following conditions:

A. *There is a normal distribution for measurements;*
B. *There is no independency (autocorrelation) between measurements*

However, in a real-world situation, project performance measurements, such as EDI and CPI values may not have a normal distribution and independency conditions. For this reason, before using control charts, we need to make sure that the project performance measurements are normally distributed and independent. Therefore, we introduce a two-step adjustment framework to make sure that measurements used by the control charts technique are independent and normally distributed. Figure 2 demonstrates the diagram of the proposed adjustment framework.

### 5. Two-step Adjustment Framework

As illustrated in Figure 2, in the first step, the normality of measurements is statistically tested by using Anderson- Darling hypothesis (D'Agostino and Stephens, 1986). This test can be performed by statistical packages such as *Minitab*. In this study, we test the normality of *EDI* and *DPI* values collected over the time. According to the normality test includes the following hypothesis:

  *H0: Measurements follow a normal distribution*

  *H1: Measurements do not follow a normal distribution*

The results of the normality test would lead us to reject or accept the null hypothesis. If the null hypothesis is not rejected, it is implied that the distribution of data is most likely normal.

Otherwise, the null hypothesis is rejected, which is confirming that the distribution is not normal.

If measurements are not distributed normally, an appropriate transformation method should be applied in order to generate a normal distribution for them. The normality of transformed results is checked again until the distribution of measurements becomes normal. In this research, *Johnson transformation* is applied in order to obtain normally distributed data.

In the second step, time series analysis is completed to remove the autocorrelation of the data. Because statistical theory shows that the existence of autocorrelation in data would result in generating false alarms when the control charts technique is used to monitor deviations. This analysis is completed by checking the autocorrelation among measurements and verifying the stationary of data. Statistical software uses different ways to check the autocorrelation of the data and depict autocorrelation graphs on different lags. If data are autocorrelated, time series models can be applied to remove the autocorrelation of the data; however, the stationarity of data must be verified to accurately apply time series methods and remove the dependency from autocorrelated data. Non-stationary data are very difficult to be modelled and forecasted as its mean value is usually changing over time.

Therefore, we define a "stationary checking" procedure that evaluates the statistical structure of measurements, resulted from step 1. If the statistical structure of measurements is not stationary, the differencing method is used to convert it to a stationary phase. Differencing method converts each element of time series data ($Y_t$) to $\acute{Y}_t$ with respect to its difference from $t-k$ element ($Y_{t-k}$) which means $\acute{Y}_t = Y_t - Y_{t-k}$, here $k$ is the autocorrelation lag. Now the stationary data is ready to feed time series models that remove the autocorrelation among measurements.

Time series analysis deals with a set of measurements $\{Y_t: t = 1, 2, ..., T\}$, which are collected at regular time intervals such as hourly, daily, monthly, or yearly. Time series methods use mathematical techniques to extract meaningful statistical information and predict the future values of measurements with respect to the sequence of previous data. This capability enables the proposed two-step adjustment framework to statistically model the correlations among measurements ($Y_t$) and remove their autocorrelation by extracting a set of residuals from data (Bisgaard & Kulahci, 2011). Then according to the framework, we will use these residuals to feed Shewhart individuals control charts.

Among different time series models, we use ARIMA (Autoregressive Integrated Moving Average) method. As discussed in (Chatfield, 2016), ARIMA is a modelling method that consists of two fragments, referred as an autoregressive part (AR) and a moving average part (MA). The autoregressive fragment involves regressing the data on its own lagged or past values. The errors resulted from the AR regressing are modelled by the moving average fragment at various points in time. With this approach, ARIMA method has a great advantage by removing any *k* lag autocorrelation among measurements, and providing much purer uncorrelated results, that are named residuals. Statistical packages such as *Minitab* fits the ARIMA model to data sets and provides us the residuals of this model. These residuals usually distributed normally; if not, applying the *Johnson transformation* method in the first step of the proposed framework provides us normally distributed residuals. Residuals are then used by individual control charts technique to monitor the project cost and time variations over the time.

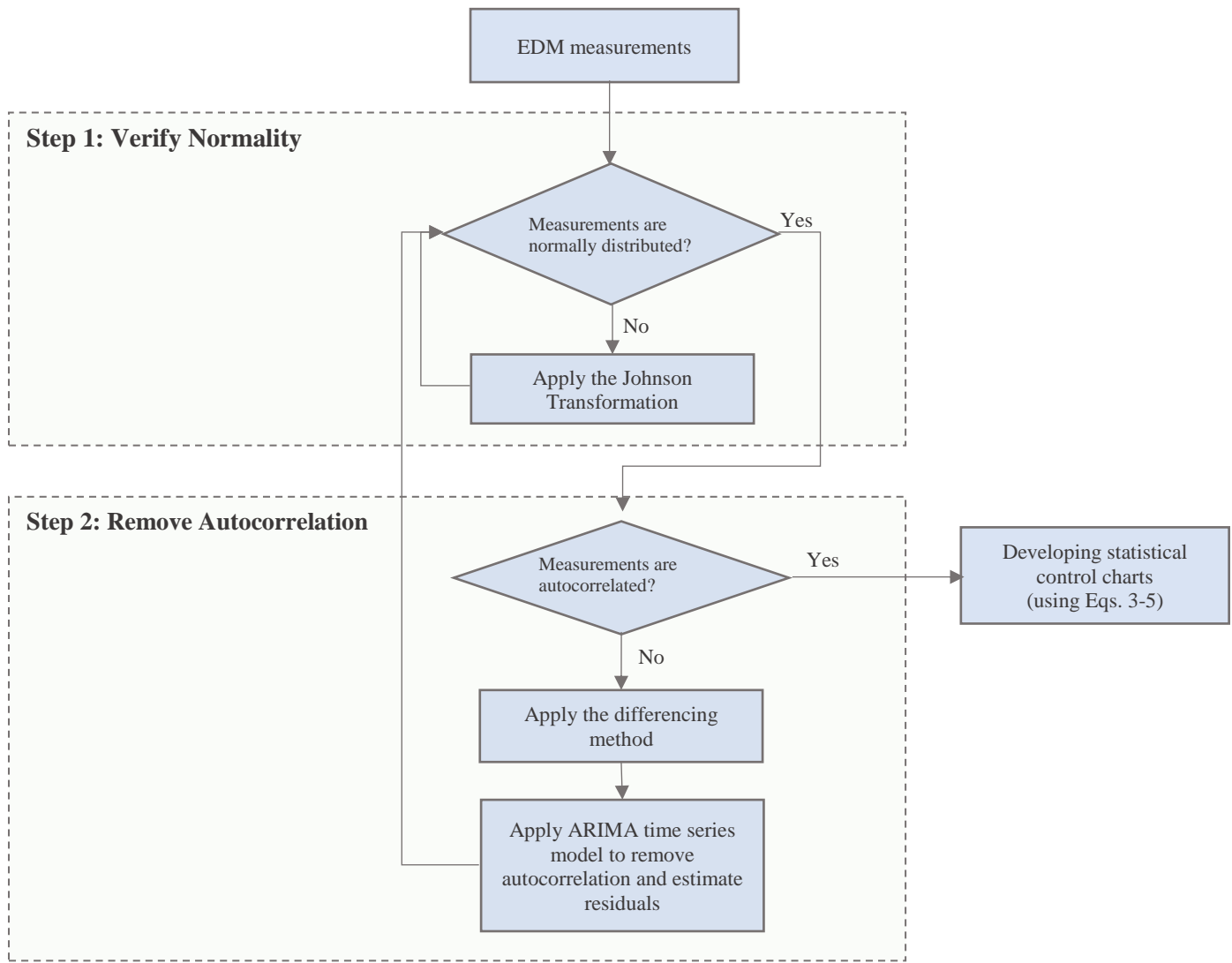

Figure 2: Two-step adjustment framework

## 6. Case study analysis

To clarify the performance of the proposed method, we evaluate the performance of a construction project by analysing EDI and DPI indices using the proposed method. Control charts will provide a more effective analysis differentiating the controlled and accepted deviations from the project plan from uncontrolled deviations that require further investigation and corrective actions. Minitab® software is used for normality test, Johnson transformation and time series analysis. The primarily data is collected weekly from the project over the first 42 weeks of the project. We computed EDI and DPI using Equations (1) and (2) and we also include the CPI as proposed in classic EVM as and efficient cost performance index. Table 1 demonstrated values of EDI, DPI and CPI in this project.

Table 1: EDM indices measurements

| Week | EDI | DPI | CPI | Week | EDI | DPI | CPI |
|---|---|---|---|---|---|---|---|
| 1 | 0.977 | 1.000 | 0.671 | 22 | 0.697 | 0.818 | 0.711 |
| 2 | 0.487 | 1.000 | 0.504 | 23 | 0.704 | 0.826 | 0.716 |
| 3 | 0.603 | 0.700 | 0.760 | 24 | 0.770 | 0.833 | 0.743 |
| 4 | 0.834 | 0.750 | 0.759 | 25 | 0.754 | 0.800 | 0.764 |
| 5 | 0.895 | 0.800 | 0.764 | 26 | 0.715 | 0.769 | 0.769 |
| 6 | 0.629 | 1.000 | 0.765 | 27 | 0.735 | 0.778 | 0.780 |
| 7 | 0.620 | 0.857 | 0.734 | 28 | 0.724 | 0.750 | 0.799 |
| 8 | 0.686 | 0.875 | 0.737 | 29 | 0.745 | 0.759 | 0.809 |
| 9 | 0.681 | 0.778 | 0.799 | 30 | 0.770 | 0.767 | 0.820 |
| 10 | 0.618 | 0.800 | 0.797 | 31 | 0.804 | 0.806 | 0.833 |
| 11 | 0.671 | 0.727 | 0.816 | 32 | 0.813 | 0.781 | 0.850 |
| 12 | 0.665 | 0.667 | 0.814 | 33 | 0.818 | 0.758 | 0.855 |
| 13 | 0.613 | 0.692 | 0.801 | 34 | 0.818 | 0.765 | 0.850 |
| 14 | 0.643 | 0.643 | 0.768 | 35 | 0.838 | 0.743 | 0.858 |
| 15 | 0.632 | 0.667 | 0.784 | 36 | 0.844 | 0.722 | 0.884 |
| 16 | 0.622 | 0.625 | 0.790 | 37 | 0.872 | 0.730 | 0.899 |
| 17 | 0.619 | 0.706 | 0.803 | 38 | 0.870 | 0.711 | 0.908 |
| 18 | 0.676 | 0.889 | 0.809 | 39 | 0.870 | 0.692 | 0.917 |

| | | | | | | |
|---|---|---|---|---|---|---|
| 19 | 0.748 | 0.842 | 0.773 | 40 | 0.874 | 0.675 | 0.930 |
| 20 | 0.733 | 0.800 | 0.741 | 41 | 0.895 | 0.683 | 0.944 |
| 21 | 0.682 | 0.762 | 0.731 | 42 | 0.893 | 0.667 | 0.959 |

6.1 Performance analysis using the proposed framework

To evaluate the cost and schedule deviations of the project from the baseline plan, EDM measurements were fed to our proposed two-step adjustment framework (Figure 2) to generate appropriate data for control charts technique. In the first step, the normality of EDM indices are evaluated individually by completing the associated hypothesis tests. Figures 3 to 5 the results of the normal distribution test for EDI, DPI and CPI measurements. Assuming the significance level at 0.20, it can be concluded from Figure 3 that EDI measurements are normally distributed, because the resulted p-value is 0.222 which greater than the level of significance. However, Figure 4 and 5 show that the small p-values of DPI lead to the rejection of the null hypothesis. Hence, CPI and DPI measurements do not have normal distributions.

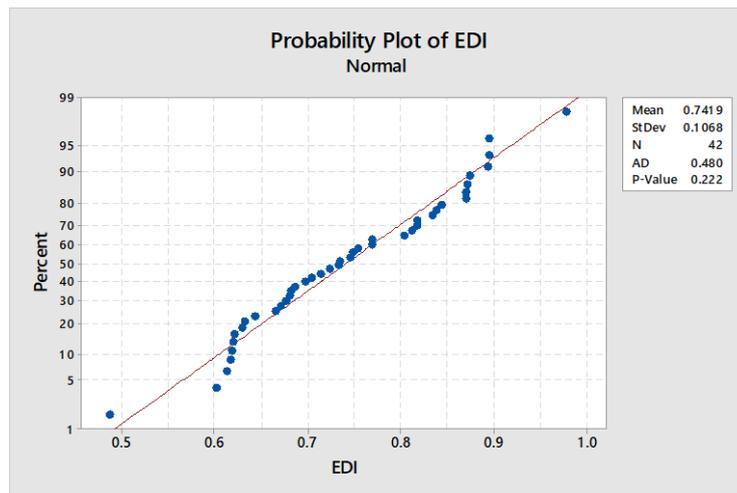

Figure 3: Checking normality of EDI

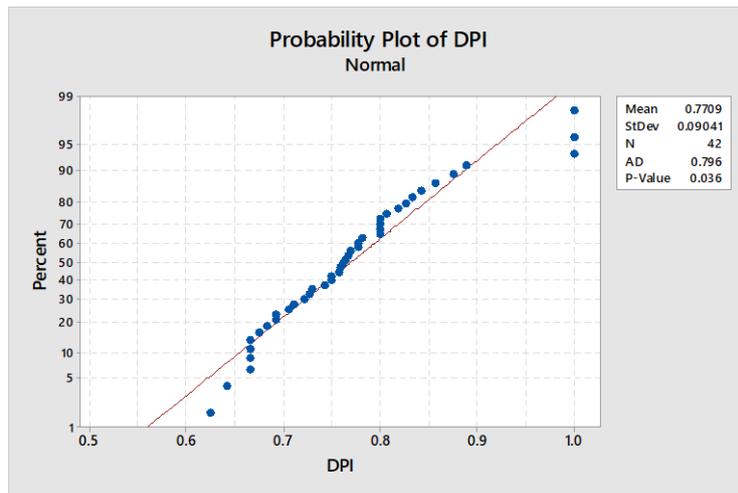

Figure. 4: Checking normality of DPI

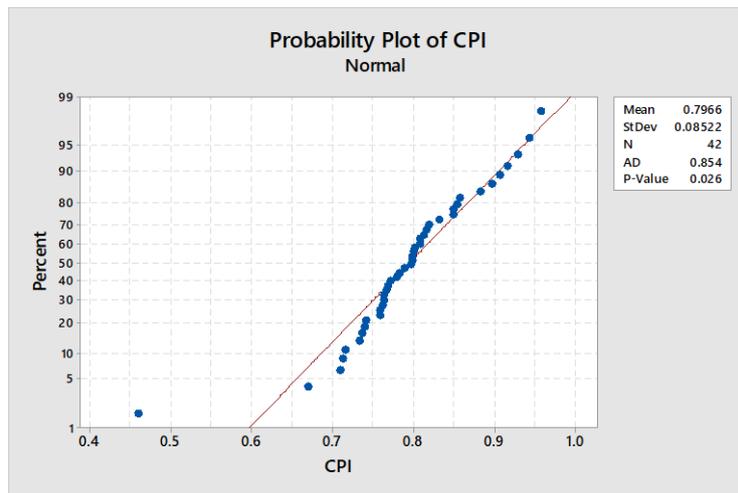

Figure 5: Checking normality of CDI

Using control charts methods for non-normal distributed data will cause inaccurate results. Hence, according to the proposed adjustment framework, DPI and CPI are individually transformed by Johnson transformation model to be normal distributed data. Table 2 shows the transformed results.

Table 2: Transformed EDM measurements to become normal

| Week | DPI transformed | CPI transformed | Week | DPI transformed | CPI transformed |
|---|---|---|---|---|---|
| 1 | -1.865 | 1.965 | 22 | -1.417 | 0.660 |
| 2 | -3.054 | 1.965 | 23 | -1.349 | 0.738 |
| 3 | -0.624 | -1.410 | 24 | -0.936 | 0.807 |
| 4 | -0.646 | -0.140 | 25 | -0.558 | 0.470 |
| 5 | -0.563 | 0.470 | 26 | -0.454 | 0.110 |
| 6 | -0.541 | 1.965 | 27 | -0.234 | 0.215 |
| 7 | -1.091 | 1.020 | 28 | 0.123 | -0.140 |
| 8 | -1.040 | 1.166 | 29 | 0.301 | -0.025 |
| 9 | 0.127 | 0.215 | 30 | 0.477 | 0.078 |
| 10 | 0.088 | 0.470 | 31 | 0.671 | 0.539 |
| 11 | 0.418 | -0.463 | 32 | 0.897 | 0.257 |
| 12 | 0.380 | -1.410 | 33 | 0.948 | -0.039 |
| 13 | 0.151 | -1.002 | 34 | 0.892 | 0.053 |
| 14 | -0.480 | -1.778 | 35 | 0.987 | -0.239 |
| 15 | -0.160 | -1.410 | 36 | 1.254 | -0.538 |
| 16 | -0.044 | -2.038 | 37 | 1.387 | -0.427 |
| 17 | 0.198 | -0.788 | 38 | 1.463 | -0.716 |
| 18 | 0.304 | 1.273 | 39 | 1.534 | -1.002 |
| 19 | -0.377 | 0.888 | 40 | 1.626 | -1.278 |
| 20 | -0.970 | 0.470 | 41 | 1.723 | -1.152 |
| 21 | -1.386 | 0.017 | 42 | 1.815 | -1.410 |

By this transformation, the new values of both DPI and CPI are normally distributed. For instance, Figure 6 shows the summary of Johnson transformation for DPI. The p-value was increased to 0.5777 when Johnson transformation was completed. Hence, the null hypothesis is accepted and transformed DPI is now normally distributed. These transformed values are our new DPI and EDI measurements used to control schedule performance variations of the project.

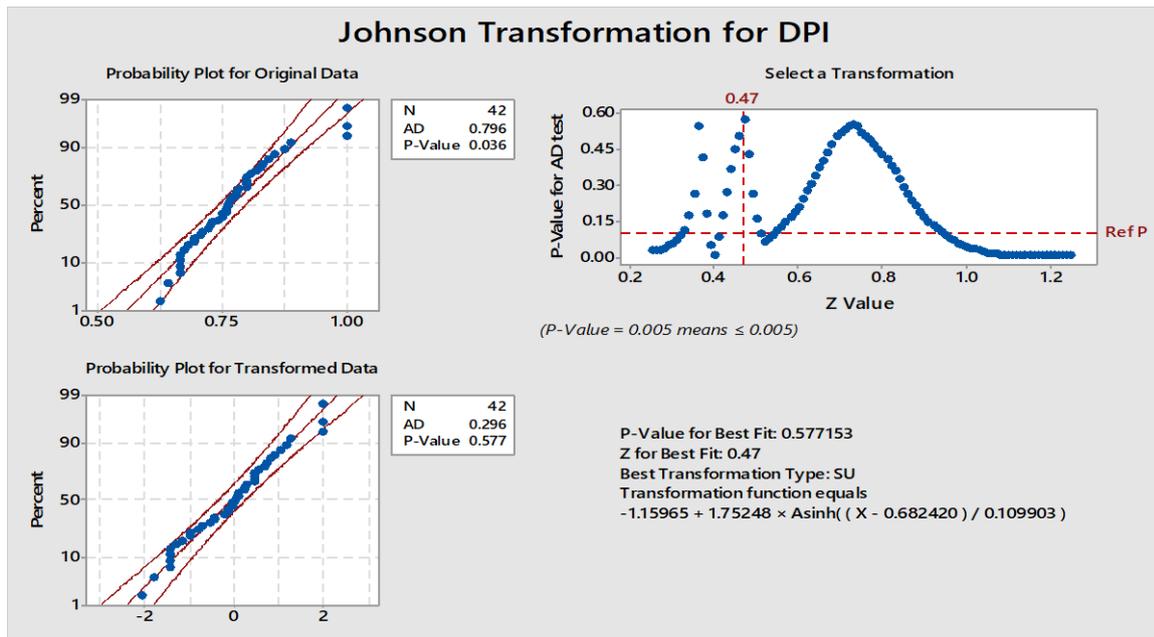

Figure 6: Summary of Johnson transformation for DPI

Now, all the three EDM measurements are normally distributed and they are passed to the second step of the proposed framework (Figure 2). In the second step, EDI, DPI, and CPI measurements were individually evaluated against their autocorrelations. According to the proposed framework, the stationary of EDM indices was checked first. Figure 7(a), 8(a), and 9(a) show the time series plots of EDI, DPI and CPI measurements.

These plots demonstrate that EDM indices are non-stationary which means their major statistical characteristics (e.g., mean, variance) are not constant during 42 weeks. Therefore, differencing method was used to convert them to stationary sets of data. According to differencing method, *k* was arbitrary set to 1 to start differencing. The resulted EDI, and DPI measures become stationary after applying this one lag differencing. Figures 7 (b), 8 (b), and 9 (b) show the time series plots of EDI, DPI proving the stationary of their measures.

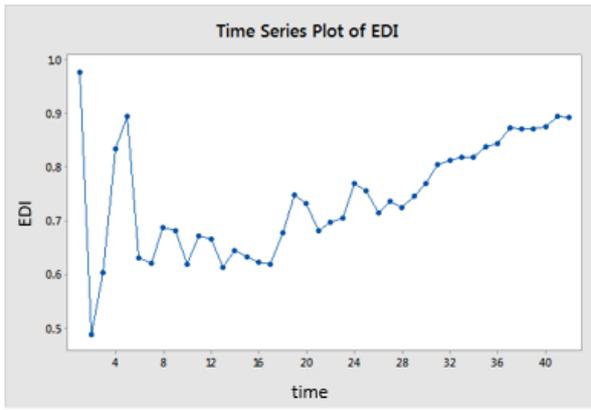 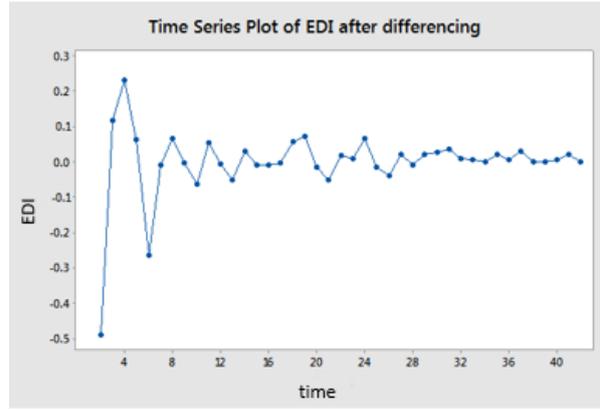

(a) (b)

Figure 7: Time series plot for EDI before and after one lag differencing

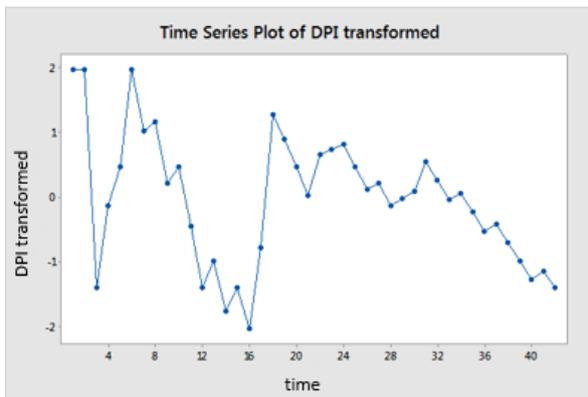 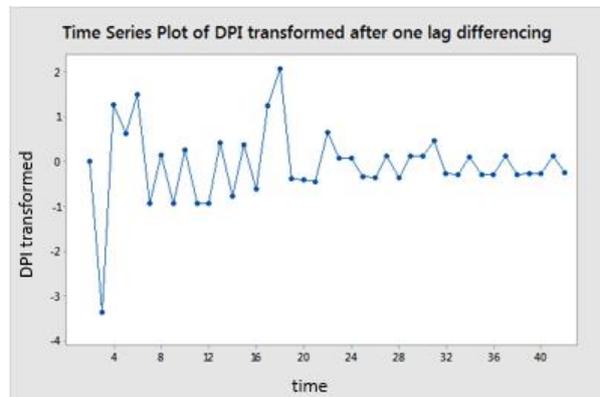

(a) (b)

Figure 8: Time series plot for DPI transformed before and after one lag differencing

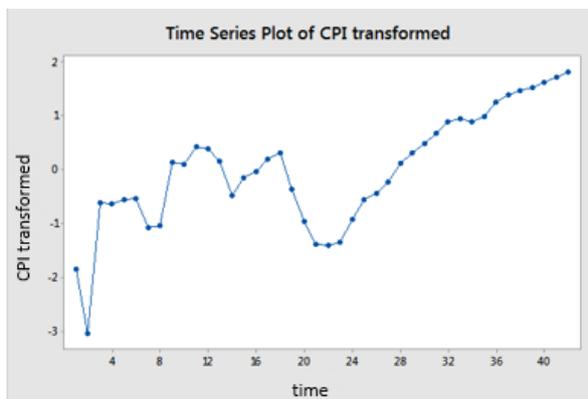 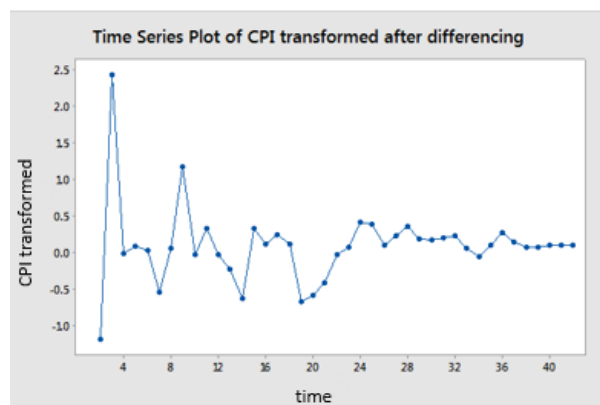

(a) (b)

Figure 9: Time series plot for CPI transformed before and after one lag differencing

In the next step, the autocorrelation among EDM measurements was checked separately. For this reason, the autocorrelation and partial autocorrelation functions are used. Autocorrelation function uncovers the possible correlation (dependence) of a variable with its own past and future values during a time. Partial autocorrelation function estimates the partial correlation of a variable with its own lagged values. According to the statistical theory (Chatfield, 2016), using both autocorrelation and partial autocorrelation functions helps practitioners to construct a more reliable ARIMA model for each index by determining the best order (kth lag) of the model (Russo, Camargo, & Fabris, 2012). Figure 10, 11, and 12 show the results of autocorrelation and partial autocorrelation plots for CPI, DPI and EDI. These figures prove that EDM indices have autocorrelation among their own values.

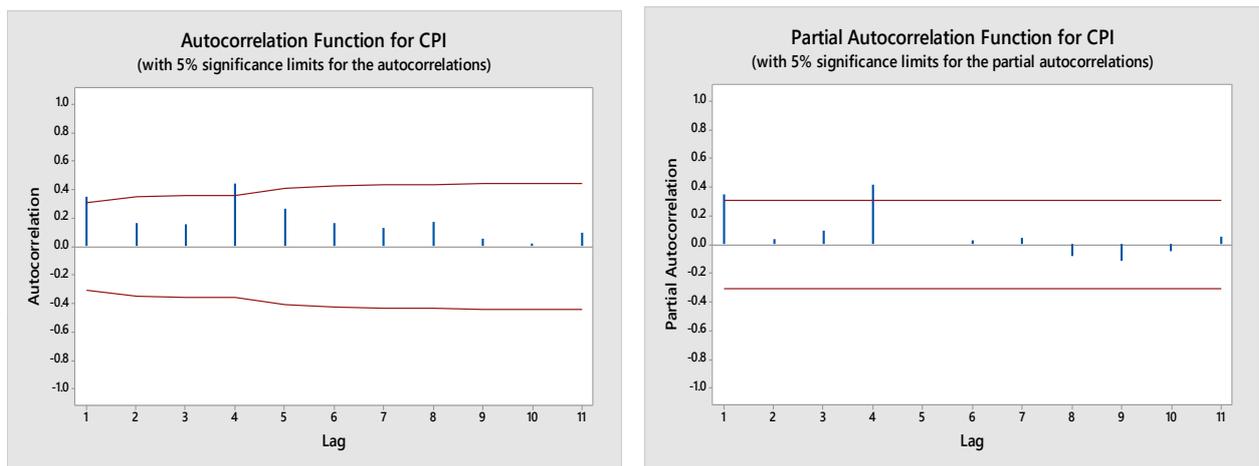

Figure 10: Autocorrelation and partial autocorrelation plots for CPI with 5% significance limits

For instance, Figure 10 shows two spikes at lag 1 and 4 on both autocorrelation and partial autocorrelation plots of CPI. These spikes conclude that each CPI measurement at time *t* is a linear function of *Y* at time *t-1* and *t-4*.

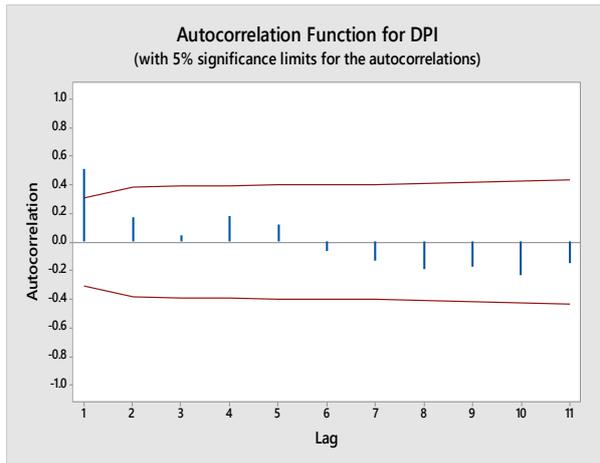 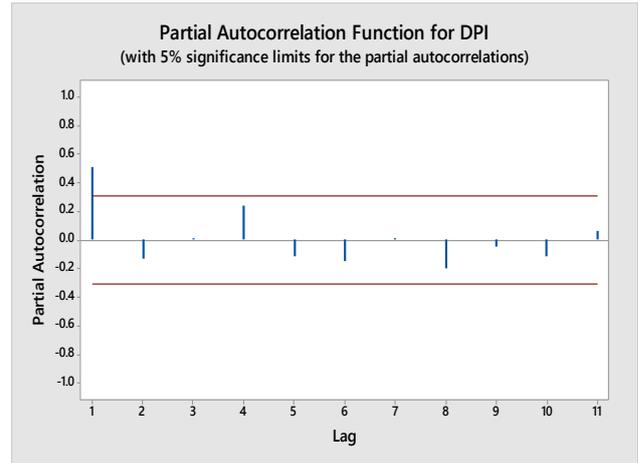

(a) *Autocorrelation of DPI*  (b) *Partial autocorrelation of DPI*

Figure 11: Autocorrelation and partial autocorrelation plots for DPI with 5% significance limits

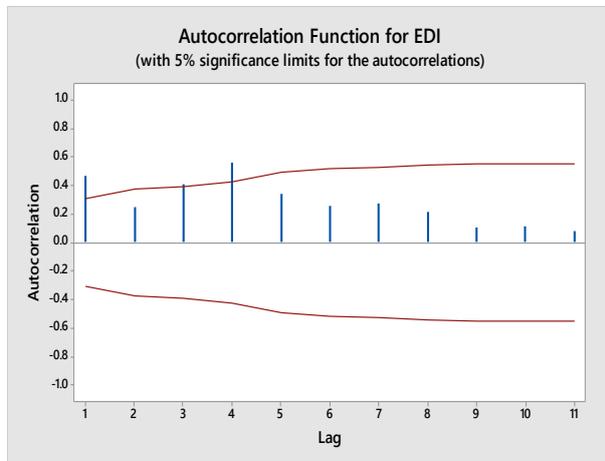 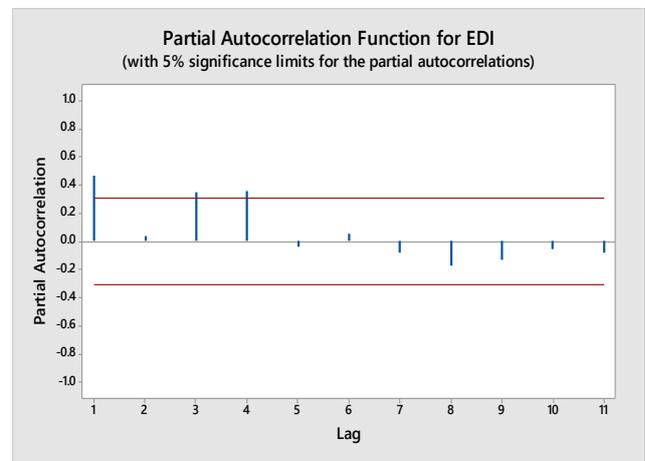

(a) *Autocorrelation of EDI*  (b) *Partial autocorrelation of EDI*

Figure 12: Autocorrelation and partial autocorrelation plots for EDI with 5% significance limits

For instance, Figure 11 demonstrates a spike in the first lag of autocorrelation and partial autocorrelation plots of DPI. These spikes that are higher than the correlation boundaries confirming that DPI measurements are auto-correlated. Hence, the first order ARIMA model can be the best time series model to detect the correlation among DPI measurements. This type of time series modelling concludes that each *Y* (measurement value) at time *t* is a linear function of *Y* at time *t-1*. Figure 12 also demonstrates three spikes on lag 1, 3, and 4 for EDI. These spikes prove that each EDI measurement (*Y*) at time *t* can be modelled as a linear function of *Y* at time *t-1*, *t-3*, and *t-4*. To determine the best values for ARIMA parameters, autocorrelation results are compared with autocorrelation characteristics suggested by Wei (1994) to set the ARIMA parameters in ways to capture the maximum correlations among measurements of each EDM indices. Our final results conclude to use ARIMA (4,1,1) for CPI, ARIMA (1,1,1) for DPI, and ARIMA (4,1,1) for EDI. These ARIMA time series models removed the autocorrelation among all EDM measurements. Therefore, their corresponding residuals are ready to develop control charts.

After checking the normality and stationarity of residuals, control charts were developed for residuals of EDM indices to monitor their deviations from the control limits. Figure 13, 14, and 15 demonstrate the results of individuals control charts technique for residuals of CPI, DPI and EDI. Control limits are calculated using Equations (3-5).

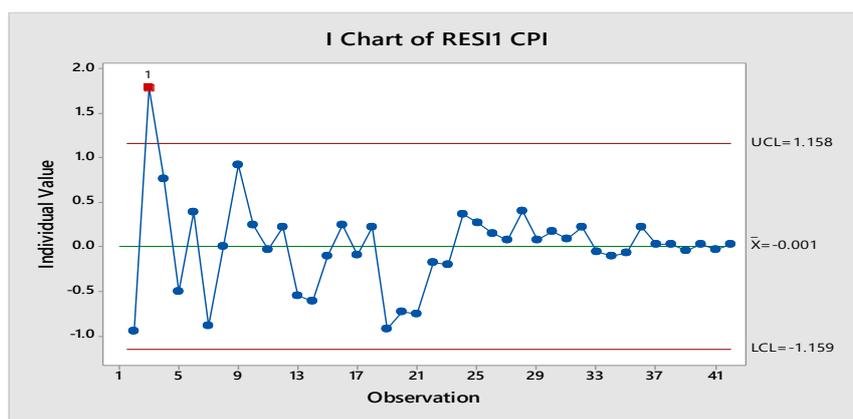

Figure 13: Special (non-random) cause control chart for CPI

As EDM indices are usually more accurate than traditional EVM indicators (Khamooshi & Golafshani, 2014), their control charts provide better insights regarding the deviations of a project from the baseline plan. Furthermore, analyzing EDM control charts and tracking the variation of EDM measurements during the project life cycle provide valuable information regarding the trends of cost and schedule performance of the project.

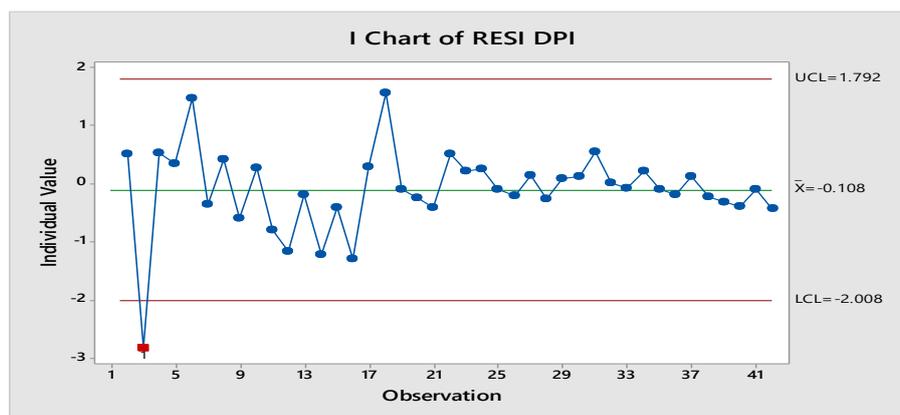

Figure 14: Special (non-random) cause control chart for DPI

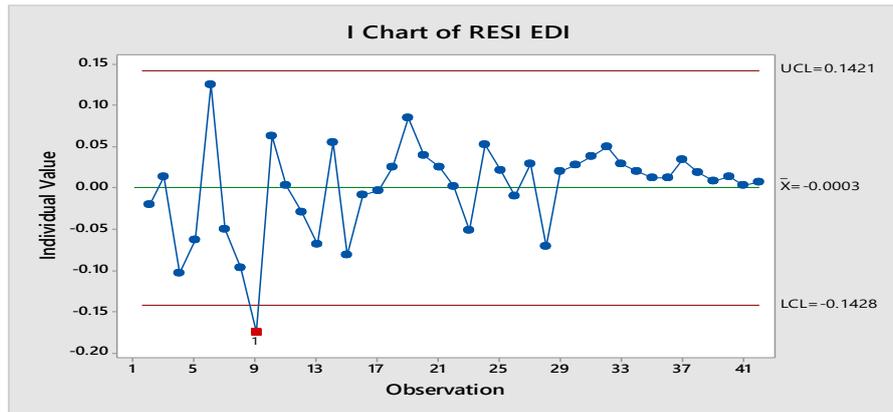

Figure 15: Special cause (non-random) control chart for EDI

For instance, according to Figure13 and 14, the schedule performance of the project has larger variations around the central control limit between week 1 and week 23. These variations become smooth after week 23 for DPI and week 29 for EDI. By reviewing the actual values (measurements) of EDI and DPI over the time (Table 1), it is implied that the project management team provided more accurate estimations regarding time (schedule) elements of the project activities after week 23. Moreover, project management team had better controls on the (on-time) completion of these activities after week 23 in compared to the first couple of weeks of the project.

According to the control charts theory, EDM measurements inside the control limits do not require immediate actions and their changes may relate to random causes that are acceptable variations. However, some EDM measurements fell out of control limits (i.e., an EDI measurement in week 9) alerting that the project performance significantly deviates from the baseline plan and project managers should find the causes of these out of control points, and do the necessary corrective actions to prevent from happening in the future. An EDM measurement above the upper control limit (UCL), at a given time $t$, shows a better performance for the project in comparison with the baseline plan. Tracing the reasons of such an enhanced performance and letting those reasons be repeated again during the project life cycle would improve the overall project performance. On the other hand, an EDM measurement below the lower control limit (LCL) shows a poor performance in comparison with the baseline plan during a given time slot. Detecting such a project performance measurement helps managers to better track associated reasons and provide proper solutions to improve the overall performance of the project. For instance, Figure 13 shows that the variation of the cost performance of the project becomes smoother in the last 6 weeks. Furthermore, there is one CPI measurement above the upper control limits related to a better cost performance in week 3. To a better construction management/ monitoring, the cause of this improvement should be detected and repeated in the future that improve the overall cost performance of the project.

Figure 14 and 15 demonstrate that most of the variation of schedule performance measurements is inside the control limits which are acceptable and they do not require immediate corrective actions. There are two out of control observations in week 3 and 9 for DPI and EDI, respectively. These out of control points alert that there were serious problems in managing the project schedule in both weeks (3 and 9), while according to the CPI control chart, the cost performance of the project was performing well during the corresponding weeks. Without using control charts, it is very difficult to detect the significant deviations of the schedule performance of the project. For example, according to DPI original values (shown in Table 1), the schedule is on-plan for weeks 1 and 2, and DPI value is equal to 0.7 for week 3, showing that the project is slightly behind the plan. This slight deviation from the baseline plan may or may not be acceptable according to the project performance expectations. But, we cannot be sure of that by only considering the original DPI measures. However, by reviewing the control chart results (Figure 14), we assure that the deviation of DPI in week 3 is not acceptable due to non-random problems and it requires an immediate investigation.

With a similar approach, we are able to detect the out of control schedule performance by reviewing EDI measurements. Table 1 shows that EDI original values are less than 1 but have not changed significantly between weeks 6 and 10. But, Figure 15 demonstrates that the residuals of EDI in week 9 is below the LCL. Therefore, the review of EDI original measures and the associated control chart implies that the project schedule is performing less than acceptable level during week 9 while the schedule deviations from the baseline plan for week 6, 7, 8, and 10 are acceptable. During weeks 3, 8 and 9 procurement activities such as delivering materials were experiencing some major delays which was the main cause of non-acceptable
In short, it is concluded that tracking the trends of original values of EDI and DPI is not sufficient for detecting significant deviations of the project performance and it is necessary to generate EDI and DPI control charts to have a better picture of the variation of the project's schedule performance. These numerical results confirm the validity of our developed approach in detecting events that result in significant delays for the project.

## 7. Conclusion

One of the most critical problems that project managers encounter is the management of their project performance, which may result in project cost and time overrun. Some unpredictable issues may occur during the duration of a project, but a proper planning and monitoring system can help managers to assuage these issues. Traditional EVM methods were developed to help managers better understand and control the project performance by determining the current status and predicting the future status of the project. Nevertheless, the traditional EVM have some deficiencies which may lead to inaccurate results in some situations. Too much emphasis on cost factors and using monetary-based indices as the main proxies for measuring the schedule performance of a project are the most important shortcomings of the traditional EVM methods. Decoupling the cost and schedule dimensions of a project can solve the problem of traditional EVM systems in calculating the schedule and duration performance of projects. In this regards, Khamooshi & Golafshani (2014) introduced the new concept of "earned duration" to focus on the duration/schedule aspect of the project. The advantages of EDM in the evaluation of the project performance have been addressed in several studies. However, EDM is not established to discriminate between acceptable and non-acceptable levels of deviations from the baseline. This study applied statistical control charts to monitor the three key indices of EDM: EDI, DPI and CPI. We proposed a two-step adjustment framework that checks the normality and dependency of EDM measurements and provides solutions for non-normally distributed and auto correlated measurements which are very common in practice. In the first adjustment step, the normality of EDM indices was analyzed and non-normal data were transformed by using the Johnson technique to be normally distributed. In the second adjustment step, the autocorrelation statistical hypothesis tests were employed to check the independency of the cost and schedule performance indices. In practice, usually CPI, DPI, and EDI indices have dependencies (autocorrelation) among their own values with different lags. Therefore, ARIMA time series model was applied for each EDM index to remove the autocorrelation among its measurements. ARIMA is a robust time series model that maximizes the reduction of the autocorrelations and provides more accurate results in compared with applying simple time series models. After removing autocorrelations, the statistical control charts technique was applied to monitor the variations of EDM indices during the project life cycle.

Our numerical findings in the case study prove that applying the statistical control charts technique provides valuable information about the project performance and its deviation from the plan. This information supports project managers to better control projects and detect unacceptable variations in the project performance that may result in unexpected cost and time losses. Furthermore, the results conclude that employing the control charts method along with analyzing the original values of EDM performance indices over the project life cycle will increase the capability of project management teams to detect cost and schedule problems on time, prevent the problems from happening in the future, and make sure about the acceptable completion of their projects.

The main purpose of EVM and EDM is to predict the future of projects, however, projects are surrounded by a variety of uncertainties which make the prediction a very difficult task. As a future research, our proposed approach can be improved by utilizing fuzzy time series to provide a more practical solutions in predicting values in situations with no trend in the previous performance.